\documentclass[prl,reprint]{revtex4-1}
\usepackage[version=3]{mhchem} 
\usepackage{units}

\begin{document}

\title[Fano Resonances in MIR Spectra of SWNTs]{Fano Resonances in Mid-Infrared Spectra of Single-Walled Carbon Nanotubes}

\author{Fran\c{c}ois Lapointe}
\affiliation{D\'{e}partement de chimie, Universit\'{e} de Montr\'{e}al, C.~P. 6128, Succursale Centre-Ville, Montr\'{e}al, Qu\'{e}bec H2C~3J7, Canada}
\affiliation{Regroupement Qu\'{e}b\'{e}cois sur les Mat\'{e}riaux de Pointe (RQMP)}

\author{\'{E}tienne Gaufr\`{e}s}
\affiliation{D\'{e}partement de chimie, Universit\'{e} de Montr\'{e}al, C.~P. 6128, Succursale Centre-Ville, Montr\'{e}al, Qu\'{e}bec H2C~3J7, Canada}
\affiliation{Regroupement Qu\'{e}b\'{e}cois sur les Mat\'{e}riaux de Pointe (RQMP)}

\author{Isabelle Tremblay}
\affiliation{D\'{e}partement de chimie, Universit\'{e} de Montr\'{e}al, C.~P. 6128, Succursale Centre-Ville, Montr\'{e}al, Qu\'{e}bec H2C~3J7, Canada}
\affiliation{Regroupement Qu\'{e}b\'{e}cois sur les Mat\'{e}riaux de Pointe (RQMP)}

\author{Nathalie Y-Wa Tang}
\affiliation{D\'{e}partement de chimie, Universit\'{e} de Montr\'{e}al, C.~P. 6128, Succursale Centre-Ville, Montr\'{e}al, Qu\'{e}bec H2C~3J7, Canada}
\affiliation{Regroupement Qu\'{e}b\'{e}cois sur les Mat\'{e}riaux de Pointe (RQMP)}

\author{Patrick Desjardins}
\affiliation{D\'{e}partement de g\'{e}nie physique, \'{E}cole Polytechnique de Montr\'{e}al, Montr\'{e}al, C.~P. 6079, Succursale Centre-ville, Montr\'{e}al, Qu\'{e}bec H3C~3A7, Canada}
\affiliation{Regroupement Qu\'{e}b\'{e}cois sur les Mat\'{e}riaux de Pointe (RQMP)}

\author{Richard Martel}
\email{r.martel@umontreal.ca}
\affiliation{D\'{e}partement de chimie, Universit\'{e} de Montr\'{e}al, C.~P. 6128, Succursale Centre-Ville, Montr\'{e}al, Qu\'{e}bec H2C~3J7, Canada}
\affiliation{Regroupement Qu\'{e}b\'{e}cois sur les Mat\'{e}riaux de Pointe (RQMP)}

\begin{abstract}
This work revisits the physics giving rise to the carbon nanotubes phonon bands in the mid-infrared. Our measurements of doped and undoped samples of single-walled carbon nanotubes in Fourier transform infrared spectroscopy show that the phonon bands exhibit an asymmetric lineshape and that their effective cross-section is enhanced upon doping. We relate these observations to electron-phonon coupling or, more specifically, to a Fano resonance phenomenon. We note that only the dopant-induced intraband continuum couples to the phonon modes and that defects induced in the sidewall increase the resonance probabilities.
\end{abstract}

\keywords{carbon nanotubes, mid-infrared spectroscopy, Fano resonance, electron-phonon interactions, phonon modes, doping}
\pacs{61.48.De, 78.64.Ch, 63.22.-m}

\maketitle

\begin{figure}[t]
\includegraphics{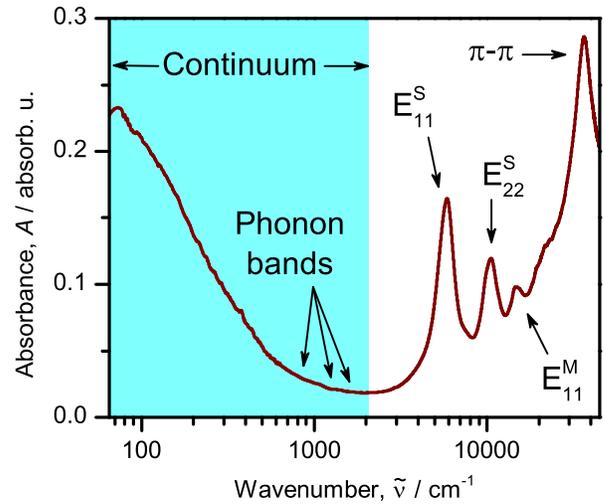}
\caption{(Color online) Absorption spectrum of a purified and annealed (undoped) SWNT film (30 nm thick). Arrows with labels indicate the regions of the continuum states, the phonon bands, and excitonic and $\pi$--$\pi$ transitions.}
\label{figsurvey}
\end{figure}

The mid-infrared (MIR) spectroscopy of carbon nanotubes has been overlooked in the past because this homonuclear allotrope of carbon does not bear permanent dipoles. Yet, a number of optical phonons are dipolar active in the infrared through the transient, and yield surprisingly rich but weak mid-infrared phonon bands~\cite{Kuhl1998,Bant2006,Berm2005,Kim2005a,Pekk2011}. As shown in Fig.~\ref{figsurvey}, these infrared active phonons are however located on the flank of a strong and broad far-infrared (FIR) electronic absorption continuum centered at around 100~cm$^{-1}$~\cite{Ugaw1999a}, which origin has been debated over the years and ascribed to both intraband free carrier excitations and interband (i.e. band-to-band) metallic transitions~\cite{Itki2002,Kama2003,Kamp2008,Akim2006}. It appears now that the coexistence of this electronic continuum absorption with the phonon bands sets favorable conditions for electron-phonon coupling in this part of the spectrum. Previous Raman and electrical transport investigations of carbon nanotubes have shown signatures of electronic coupling with optical phonons. A Breit-Wigner-Fano profile was initially used to explain the asymmetric metallic SWNTs G-band in Raman spectroscopy~\cite{Brow2001}, but this broadening was later ascribed to a Kohn anomaly~\cite{Lazz2006,Pisc2007}. Electron-phonon interactions have also been extensively explored in theory~\cite{Pere2005b} and in coupled transport-Raman experiments~\cite{Das2007} in order to explain the origin of a saturation of the drift velocity of carriers (and current) in carbon nanotubes. Nonetheless, they are still unreported in infrared absorption spectra, probably missed due to the presence of the strong FIR electronic continuum.

Herein, we present the first report of Fano resonances in the MIR spectra of single-walled carbon nanotubes (SWNTs). Our study shows that the resonance induces an asymmetric lineshape that originates from the coupling of infrared active phonon modes with the continuum of free carrier states (intraband transitions). The resonance cross-section is strongly enhanced by doping, which explains the stronger than expected infrared activity of the carbon nanotube phonons. Inducing defects on the nanotube sidewalls is found to further increase the signal of the Fano resonance. 

\begin{figure}[t]
\includegraphics{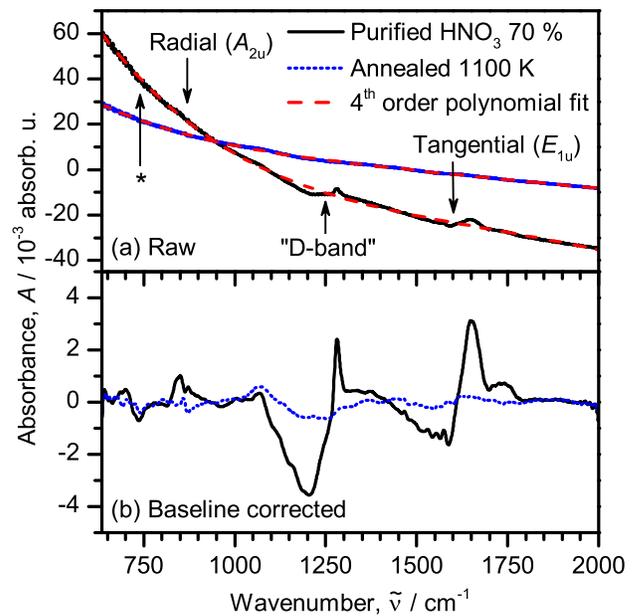}
\caption{(Color online) MIR spectra of SWNTs purified and annealed. SWNTs purified with \ce{HNO3}~70\% show asymmetric kinks at calculated phonon energies that disappear upon annealing at 1100~K under high vacuum. The symmetries indicated for reference are those of graphite ($A_\textrm{2u}$ and $E_\textrm{1u}$). Radial and tangential refer to the actual SWNTs modes. The mode at 740~cm$^{-1}$ is denoted by a star (*). (a)~Raw spectra of SWNTs showing the 4th order polynomial fit used as a baseline. (b)~Spectra after baseline subtraction.}
\label{figpurif}
\end{figure}

For our experiments, transmission spectra in the MIR of $\sim\,$30~nm thick mats of purified SWNTs (laser ablation, refluxed in concentrated nitric acid 70\%) deposited on intrinsic silicon substrates were acquired using a Br\"{u}ker Vertex 80v Fourier transform spectrometer (MCT detector, Globar source and broadband KBr beamsplitter, resolution 4~cm$^{-1}$, 1024~scans, interference fringes removed post-acquisition). These conditions were set to achieve the highest sensitivity in the relevant phonon bands region shown in Fig.~\ref{figsurvey}. The continuous black curve in Fig.~\ref{figpurif} shows what is best described as kinks ---~at 1600, 1250, 870 and 740~cm$^{-1}$~--- appearing over the rising background of the free carrier absorption in the far-infrared. However, peaks in the absorption spectra would be expected instead of the observed dips. All of these bands have been previously reported, and correspond to calculated phonon modes, with the exception of the 740~cm$^{-1}$, which origin is yet unknown to us. The kink at $\sim\,$1600~cm$^{-1}$ is attributed to symmetries $E_{1}$ for chiral, and $E_\textrm{1u}$ for zig-zag and armchair nanotubes~\cite{Eklu1995,Doba2003,Jeon2005}. This is similar to the 1590~cm$^{-1}$ $E_\textrm{1u}$ tangential mode of graphite. Additionally, the band at $\sim\,$870~cm$^{-1}$ belongs to symmetries $A$ or $E_{1}$ for chiral, $A_\textrm{2u}$ for zig-zag and $E_\textrm{1u}$ for armchair nanotubes. It is a radial mode, analogous to the 868~cm$^{-1}$ out-of-plane transverse optic (oTO) $A_\textrm{2u}$ graphite mode. Finally, the band at $\sim\,$1250~cm$^{-1}$ is associated to disorder-induced transitions and will be named hereafter the ``D-band''~\cite{Berm2005,Bant2006}. Nevertheless, the asymmetric lineshape common to all vibrational modes remained unnoticed in previous reports. Moreover, upon annealing for one hour under high vacuum (10$^{-6}$~Torr) at 1100~K, all kinks disappear, but the one at $\sim\,$870~cm$^{-1}$, leaving a smooth rising background (Fig.~\ref{figpurif}, dotted blue curve). This dependence of the phonon band intensities on the doping state of the SWNTs is unexpected, and to our knowledge has never been reported in previous studies. 

\begin{figure}[t]
\includegraphics{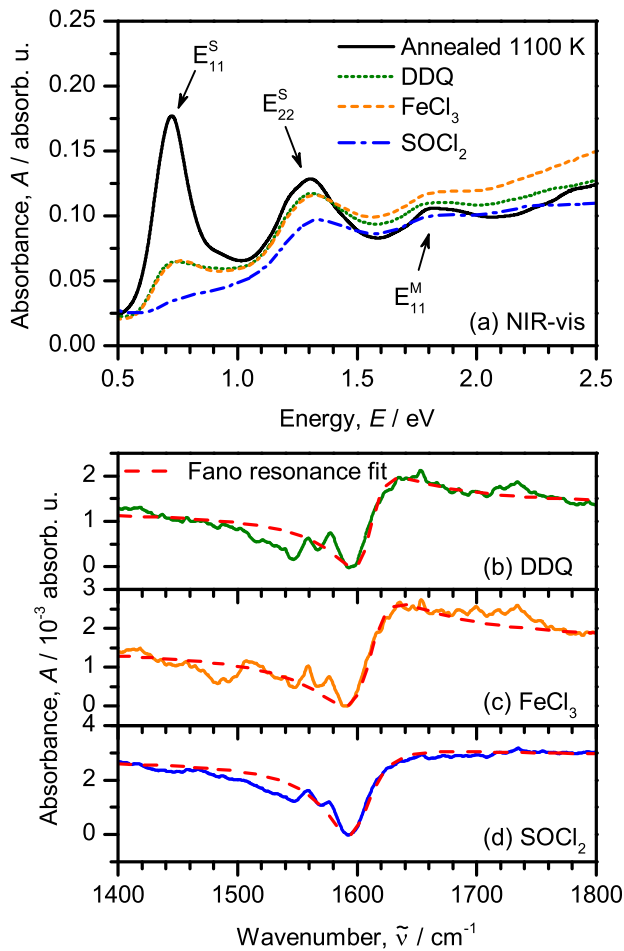}
\caption{(Color online) (a) NIR-vis transmission spectra exhibiting the E$_\textrm{ii}$ excitonic transitions of annealed (1100~K, high vacuum) and doped SWNTs with three different oxidizers. Baseline corrected MIR spectra showing the 1600~cm$^{-1}$ mode of doped SWNTs with (b) [DDQ]$=25$~mM, (c) [\ce{FeCl3}]$=25$~mM and (d) pure \ce{SOCl2}.}
\label{figdoped}
\end{figure}

To prove that the kinks do not pertain to molecular vibrational states of doping adsorbates, new films made of purified SWNTs (reflux in \ce{HNO3}~70\%) deposited on intrinsic silicon and quartz substrates were annealed for one hour at 1100~K in 10$^{-6}$~Torr vacuum and then chemically doped using three different oxidizers: 2,3-dichloro-5,6-dicyano-1,4-benzoquinone (DDQ, 25~mM in acetonitrile), \ce{FeCl3} (25~mM in acetonitrile) and \ce{SOCl2} (pure). These dopants were chosen because of their distinct chemical structures. The SWNT films were immersed in the dopant solution for two days and the resulting \emph{p}-doped layers were asserted by near-infrared--visible (NIR-vis) transmission spectroscopy (InGaAs and Si diode detectors, tungsten source, \ce{CaF2} beamsplitter, resolution 16~cm$^{-1}$, 1024~scans): In Fig.~\ref{figdoped}(a), we observe that the first transition of doped semiconductor SWNTs (E$^\textrm{S}_{11}$, $\sim\,$0.7~eV) is suppressed in comparison to the annealed SWNTs. Relative to the other two dopants, the E$^\textrm{S}_{11}$ of \ce{SOCl2}-doped SWNTs is attenuated further by the strong oxidizing conditions. Such a change in the optical spectra is due to charge transfer doping and Pauli exclusion.  Doping the SWNT films results however in the recovery of the characteristic dips (Fig.~\ref{figdoped}(b-d) and Figure~S5 in Suppl.~Mater.~\footnote{See Supplemental Material at [insert URL]Êfor wide range spectra.}) seen in the MIR spectrum of purified SWNTs (Fig.~\ref{figpurif}, continuous black curve), regardless of the type of dopant. Indeed, the MIR spectra show the same features when doped with the three different oxidizers, thus proving that these features are not of molecular origin. 

The lineshape of the kinks is characteristic of the well-known Fano resonance, which stems generally from the scattering of a continuum states onto a discrete state~\cite{Fano1961}. It is dependent on the strength of the coupling. Briefly, the scattered wavefunction suffers a $\pi$ phase shift as it goes through an interval of energies $\sim \left|V_\textrm{E}\right|^2$ around a resonance, thus yielding destructive interferences (anti-resonance), and creating an asymmetric profile in the optical density neighboring the resonance. The absorption spectra are in fair agreement with a fit of the Fano model (dashed red line, Fig.~\ref{figdoped}(b), (c) and (d)):
\begin{equation}
A(E) = \frac{a \left( q + \zeta \right)^2}{1 + \zeta^2}, \label{eqn:fano}
\end{equation}
where $A$ is the absorbance, $\zeta = 2 \left( E - E_\textrm{0} \right)/\Gamma$ is the reduced energy parameter, $a$ is an adjustment factor, $E$ is the energy, $E_\textrm{0}$ is the effective energy of the ``mixed'' discrete state shifted by a factor $F(E)$ with respect to the true position $E_\varphi$ of the discrete state, $\Gamma = 2 \pi \left| V_\textrm{E} \right|^2$ is the spectral width of the discrete state with $V_\textrm{E}$ as the matrix of the coupling elements, and $q$ is the lineshape parameter, defined as:
\begin{equation}
q = \frac{\mu_\varphi + P \int \mathrm{d}E' V_\textrm{E'} \mu_\textrm{E'} / \left( E-E' \right)}{\pi V_\textrm{E} \mu_\textrm{E}}. \label{eqn:qparam}
\end{equation}
The transition moments $\mu_\varphi$ and $\mu_\textrm{E}$ belong to the discrete state and the continuum respectively, while $P$ designates the ``principal part of'' the integral. 

We allowed the parameters $a$, $E_\textrm{0}$, $\Gamma$ and $q$ to vary during the fitting process. The fitting was performed using a limited range of the absorbance data around the best defined mode after baseline subtraction (4th order polynomial). The results are shown in Table~\ref{tbl:param}. Due to the Fano resonance, the 1600~cm$^{-1}$ mode undergoes an upshift compared to the tangential mode of graphite (1590~cm$^{-1}$)~\cite{Bant2006}. This is only an apparent shift as $E_\textrm{0} = E_\varphi + F(E)$, where $E_\varphi$ is the true position of the vibrational state. 

\begin{table}
  \caption{Fitting parameters of the Fano resonance model on MIR transmission spectra of doped SWNTs with three different oxidizers (1600~cm$^{-1}$ mode region).}
  \label{tbl:param}
  \begin{tabular}{l|ccc}
    \hline
    & DDQ  & \ce{FeCl3} & \ce{SOCl2} \\
    \hline
    $E_\textrm{0}$ / cm$^{-1}$   & 1609.5 & 1608.7 & 1599.8     \\
    $\Gamma$ / cm$^{-1}$         & 37.6   & 47     & 48         \\
    $q$                          & 0.71   & 0.81   & 0.28       \\
    $a$ / $10^{-5}\,$a. u.       & 129.0  & 158.0  & 282        \\
    \hline
    Reduced $\chi^2$ / $10^{-9}$ & 9.26   & 13.1   & 26.8       \\
    \hline
  \end{tabular}
\end{table}

Strong electron-phonon interactions also produce broader transitions than expected for pure phonon modes. Indeed, a linewidth $\Gamma$ of 38-48~cm$^{-1}$ is observed. The lifetime of the transition $\hbar / \Gamma$, according to the fit, is of the order of $\sim\,$100~fsec. In comparison, the linewidth of the longitudinal (LO) and transverse (TO) optical modes of semiconductor SWNTs in Raman spectroscopy are calculated to be $\sim\,$10~cm$^{-1}$~\cite{Lazz2006}. 

The parameter $\nicefrac{1}{2} \pi q^2$ is related to the ``ratio of the transition probabilities to the `modified' discrete state and to a band width $\Gamma$ of unperturbed continuum states''~\cite{Fano1961}. That is, $\nicefrac{1}{2} \pi q^2$ will tend to infinity if the transition probability to the discrete state dominates that to the continuum states or will tend to zero in the reciprocal case. Values in Table~\ref{tbl:param} give a parameter $\nicefrac{1}{2} \pi q^2$ of around 0.8, 1.0 and 0.1 for the DDQ, \ce{FeCl3} and \ce{SOCl2} films, respectively. We can thus conclude that the dip observed for the \ce{SOCl2} doped film in Fig.~\ref{figdoped}(d) originates mostly from the electronic states, while the kinks of the DDQ and \ce{FeCl3} doped samples include a higher contribution of the modified tangential phonon mode.

From the trend of the $q$ parameter in Eq.~\ref{eqn:qparam} upon doping, we can infer that the transition moment of the electronic continuum states $\mu_\textrm{E}$ quickly dominates the phonon transition moment $\mu_\varphi$. Moreover, the absence of infrared features in the undoped sample in Fig.~\ref{figpurif} indicates that $\mu_\varphi$ is very weak (below the sensitivity of our setup), which is expected because SWNTs have no permanent dipole. For example, the IR active mode at $\sim\,$1600~cm$^{-1}$ in Fig.~\ref{figpurif} completely vanishes upon annealing (dedoping). 

The band at $\sim\,$870~cm$^{-1}$ in Fig.~\ref{figpurif}, ascribed to the radial mode, remains observable at all times with our setup, yet it may change shape when the sample is annealed. Although the transition moment of this phonon mode can in principle explain this permanent IR activity, the shape of this band rather indicates the presence of a resonance. 
We speculate that a residual doping is left even after annealing, which is enough for the free carrier continuum to couple with the low frequency phonon modes to produce a Fano resonance.  

\begin{figure}[t]
\includegraphics{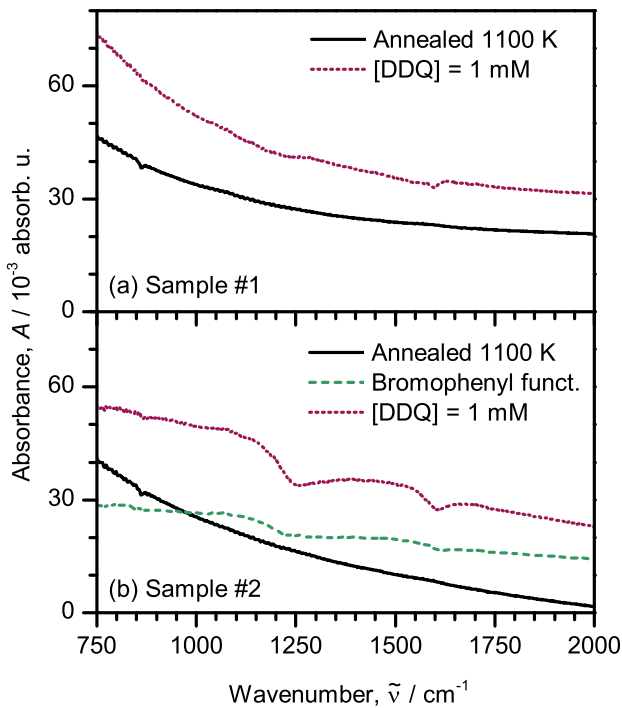}
\caption{(Color online) MIR transmission spectra on the same scale showing the effect of inducing defects in the sidewall of SWNTs. (a)~Sample~\#1. Comparison of the sample in annealed and doped state ([DDQ]$=1$~mM). (b)~Sample~\#2. Comparison of the sample in annealed state, functionalized with bromophenyls and then doped ([DDQ]$=1$~mM). Once functionalized and doped, sample~\#2 shows more acute features than doped sample~\#1.}
\label{figfunct}
\end{figure}

In order to explain the presence of the Fano resonance and the increase of its cross-section upon \emph{p} doping, it must be considered that the initial state of the transition is the bare electronic continuum, and that the final state then is the superposition of the unperturbed electronic continuum eigenstate and the discrete phonon state modified by the electron-phonon interactions.
Thus, only two levers can change the shape and intensity of the Fano resonance upon doping: The electronic and vibrational transition moments (e.g. Eq.~\ref{eqn:qparam}). As discussed before, the broad FIR absorption continuum associated to dopant-induced intraband transitions is clearly link to the lever $\mu_\textrm{E}$ at the origin of the Fano resonance. For undoped carbon nanotubes (Fig.~\ref{figsurvey}), vertical intraband transitions are forbidden because of the 1D bandstructure while interband metallic transitions dominate. The absence of Fano resonance in the undoped samples therefore implies that the interband metallic continuum does not couple to the localized phonon modes. Hence, only the intraband continuum in doped samples is active in the Fano resonance. It is however difficult to understand how such an intraband continuum can exist in the context of a strictly 1D bandstructure such as for doped carbon nanotubes. Indeed, to account for a FIR active intraband continuum, one should consider that the dopants introduce weak disorder to the nanotube bands so that the system adopts the so-called ``dirty regime'', as in other semiconductors.\cite{Blase2009} In this regime, intraband vertical transitions become allowed due to disorder induced by the presence of dopants and ionized gap states nearby the conduction or valence bands. The extent of the continuum in the FIR and its overlap with the higher energy phonon modes will thus depend on the level of disorder and hence the dopant concentration. In our \emph{p}-type doped samples, an increase of dopants leads to an accumulation of ionized states localized in the gap nearby the valence band, which broadens the nanotube bands and expands the absorption continuum toward the higher energy part of the FIR spectrum where discrete phonon modes are located. 

As described by the Fano physics, only IR active modes possess a finite vibrational transition moment $\mu_\varphi$ that enables the Fano resonance. However, the band at $\sim\,$1250~cm$^{-1}$ in Fig.~\ref{figpurif}, related to the so-called ``D-band'', is ``forbidden'' in the infrared. This D-mode is however known to be activated by the presence of impurities or defects that breaks the local symmetry.\cite{Berm2006} In this context, Bermudez showed that most of the MIR phonon mode intensities depend on the number of defects in the sidewall of SWNTs~\cite{Berm2005,Berm2006}.  
We thus verify qualitatively the dependence of the Fano resonances in the MIR on the number of defects.
Using the diazonium radical reaction to graft bromophenyls to the sidewalls, we compare side by side the MIR spectra of annealed and functionalized SWNTs (Fig.~\ref{figfunct}), both before and after doping with DDQ (1~mM in acetonitrile, 30~min.)
For the reaction, a solution at a concentration of 0.79~mM of 4-bromobenzenediazonium tetrafluoroborate (96\%, Sigma-Aldrich) was prepared with degassed deionized water (Millipore Milli-Q, 18.2~M$\Omega$) and the pH was adjusted to $\sim\,$10 with NaOH \cite{Dyke2003}. 
The SWNT film was dipped into the aqueous salt solution for 10~minutes, rinsed with water, then with 2-propanol (Certified ACS Plus, Fisher Scientific) and finally dried using a \ce{N2} stream. 
Raman measurements were used to monitor the reaction (see Fig.~S3 and Fig.~S4 in Suppl. Mater.~\footnote{See Supplemental Material at $[insert URL]$~ for Raman spectra.})
The film was subsequently doped by immersion in a 1~mM DDQ solution in acetonitrile for 30~min. 
As seen in Fig.~\ref{figfunct}(b), functionalized SWNTs (dashed green curve) show weak features at the wavenumbers of the phonon modes, while the spectra of annealed samples (continuous black curves) are smooth. 
However, functionalized SWNTs present more acute bands after doping (Fig.~\ref{figfunct}(b), dotted purple curve) than unmodified SWNTs (Fig.~\ref{figfunct}(a), dotted purple curve).

Which disorder between structural defects and dopants influences the most the intensity of the Fano process, here set by the adjustment factor $a$? In the intrinsic state, the only mode observed is the $\sim\,$870~cm$^{-1}$, which suggests that most of the Fano amplitude is gained through an increase of the electronic continuum absorption by doping. This chemical doping is sufficient to enhance the resonance process by increasing the overlap between the continuum and discrete states.  At high defect concentration, induced here by covalent bonding, the number of defect sites increases, which brings further disorder and higher localized gap state density.~\cite{Zhao2004b} The increase of the adjustment factor $a$ in this case implies that there is an increase of the number of scattering centers, thereby enhancing the activity of the Fano resonance. 

Our findings about the Fano resonance can be used to quantify the dopant concentration and the defect density for a given sample of carbon nanotubes. Indeed, the transmission MIR spectroscopy of the phonon bands yields a signal that is proportional to the concentration of dopants and scattering sites, and does not seem to discriminate between SWNTs (in contrast to the case of resonance Raman spectroscopy~\cite{Dres2005}). For instance, one can use the Fano fitting parameters of the MIR D-band to quantitatively determine the level of doping and functionalization of a given sample. Considering the wide availability of Fourier transform IR spectrometers, MIR spectroscopy could therefore prove itself as a powerful probe of chemical reactions on carbon nanotubes. Yet, the main unknown from our study is the determination of the relative contributions of the metallic and semiconducting SWNTs. This calls for further experiments on enriched SWNT samples.

Our work is similar to the Fano resonance in bilayer graphene recently reported by Tang \emph{et al.}~\cite{Tang2010} and Kuzmenko \emph{et al.}~\cite{Kuzm2009}, in which case the continuum of states was ascribed to interband transitions. The propensity of graphitic nanostructures to exhibit continuum states at low energy, involving intra- or inter-band transitions, along with the available phonons inherent to carbon atom lattices, provides the essential elements for electron-phonon coupling. The Fano resonance then appears as the natural consequence of the coexistence of a highly conjugated $\pi$ system along with a primitive crystalline cell of two non-equivalent carbon atoms. 

\begin{acknowledgements}
The authors thank Benoit Simard (NRC Steacie Institute for Molecular Sciences) for providing the laser ablation SWNTs. Valentin Popov, Francesco Mauri, Delphine Bouilly, Tony Heinz and Thomas Szkopek are acknowledged for insightful discussions. 
\end{acknowledgements}

\bibliography{irfanocnt_PRL_v4}

\end{document}